\documentclass[12pt]{iopart}
\usepackage{graphicx} 

\begin{document}

\title[Quantum controlled measurement]{Sequential measurements of non-commuting observables with quantum controlled interactions}

\author{Holger F Hofmann}

\address{Graduate School of Advanced Sciences of Matter, Hiroshima University,
Kagamiyama 1-3-1, Higashi Hiroshima 739-8530, Japan} 
\address{JST, CREST, Sanbancho 5, Chiyoda-ku, Tokyo 102-0075, Japan}
\ead{hofmann@hiroshima-u.ac.jp}

\begin{abstract}
The origin of non-classical correlations is difficult to identify since the uncertainty principle requires that information obtained about one observable invariably results in the disturbance of any other non-commuting observable. Here, this problem is addressed by investigating the uncertainty trade-off between measurement errors and disturbance for measurement interactions controlled by the state of a single qubit, where the measurement is described by a quantum coherent superposition of a fully projective measurement and the identity operation. It is shown that the measurement statistics obtained from a quantum controlled measurement of $\hat{A}$ followed by a projective measurement of $\hat{B}$ can be explained in terms of a simple combination of resolution and back-action errors acting on an intrinsic joint probability of the non-commuting observables defined by the input state of the system. These intrinsic joint probabilities are consistent with the complex-valued joint probabilities recently observed in weak measurements of quantum systems and provide direct evidence of non-commutativity in the form of imaginary correlations between the non-commuting operators. In quantum controlled measurements, these imaginary correlations can be converted into well-defined contributions to the real measurement statistics, allowing a direct experimental observation of the less intuitive aspects of quantum theory.  
\end{abstract}


\maketitle
\section{Introduction}
According to textbook quantum mechanics, the precise measurement of a physical property instantaneously changes the state of the system to an eigenstate of that property. This ``collapse'' of the state eliminates all quantum coherences between alternative measurement outcomes, resulting in a change of statistics for all physical properties that do not commute with the initial measurement. It is therefore difficult to obtain direct evidence for the statistical correlations between non-commuting properties using standard projective measurements. 

Recently, weak measurements have been widely used to overcome this limitation of conventional quantum measurements \cite{Res04,Mir07,Wil08,Lun09,Yok09,Gog11,Suz12,Roz12}. In weak measurements, the trade-off between measurement resolution and back-action disturbance is used to minimize the disturbance while obtaining a very noisy measurement outcome \cite{Aha88}. Since the disturbance of the initial state is minimal, the weak value of an observable $\hat{A}$ obtained when post-selecting a final measurement outcome $b$ of a non-commuting observable $\hat{B}$ can be used to identify correlations and even joint probabilities of $\hat{A}$ and $\hat{B}$. One of the merits of this identification of non-classical correlations is that it permits a more detailed analysis of quantum paradoxes in terms non-positive probabilities \cite{Res04,Wil08,Lun09,Yok09,Gog11,Suz12}. Specifically, the joint probabilities reconstructed from weak measurement results are generally complex and correspond to quasi-probability distributions that were already derived in the early days of quantum mechanics as quantum analogs of classical phase space statistics \cite{McCoy32,Kir33,Dir45}. Weak measurements thus give direct experimental access to the non-classical features inherent in the Hilbert space algebra of the quantum formalism. Moreover, the complex joint probabilities obtained from any pair of properties with mutually overlapping eigenstates also provide a complete description of the quantum state. It is therefore possible to perform complete quantum state tomography using weak measurements of projectors on eigenstates $\mid a \rangle$ of $\hat{A}$ followed by projective measurements of the eigenstates $\mid b \rangle$ of $\hat{B}$ \cite{Joh07,Lun12,Hof12,Sal13,Bam13}. 

Despite their successful application to numerous problems in quantum physics, the relevance of weak measurement remains somewhat controversial, mainly because the weak measurement limit is an extreme case that seems to be fundamentally different from the conventional projective measurements and appears to result in values different from the ones observed directly. It is therefore important to understand how the results of weak measurements relate to those obtained by stronger measurements \cite{Kof12,Nak12,Dre12}. Unfortunately, a detailed analysis of resolution errors and measurement back-action in intermediate steength measurements is itself a rather demanding task, as shown by the weak measurement results recently obtained to identify the disturbance and the errors in Ozawa's theory of measurement uncertainties \cite{Roz12,Lun10}. To clarify the validity of weak measurement results and the physical meaning of the complex probabilities derived from them, it is therefore desirable to design measurements that can bridge the gap between the weak measurement limit and strong projective measurements with resolution errors and back-action disturbances that are sufficiently simple to retrieve the intrinsic correlations between non-commuting properties from the experimental evidence with only a minimal set of assumptions. 

In this paper, it is shown that a particularly simple trade-off between measurement back-action and resolution can be achieved in a quantum controlled measurement of a $d$-level system, allowing an unambiguous separation of measurement noise from the intrinsic probabilities of the observables. Essentially, a quantum controlled measurement uses a control qubit to implement a coherent quantum superposition of zero interaction and fully projective measurement. The measurement errors are then limited to only two distinct patterns: the resolution error given by a completely random selection of the measurement outcome, and the measurement back-action, given by the conditional probability distributions defined by the projection of the initial state onto the eigenstates of the different outcomes. The joint probabilities for the outcomes of a quantum controlled measurement of $\hat{A}$ followed by a precise measurement of $\hat{B}$ can then be separated into a sum of contributions associated with a fully projective measurement, a back-action free assignment of a random measurement result, and a third contribution that originates from quantum coherence between the two operations and appears to describe a precise measurement  without any back-action disturbance. Although the contributions of this coherence term to the experimentally observed joint probabilities are obviously real, the intrinsic joint probability that describes the statistics of this contribution depends on the phase of the control qubit, requiring the assumption of two components in the measurement independent joint probability defined by the quantum state. Thus the phase degree of freedom in the quantum controlled measurement motivates an interpretation of the quantum state as a complex joint probability. As the following discussion will show, such complex joint probabilities provide a consistent representation of the non-classical features described by the algebra of non-commuting operators in the conventional mathematical formalism of quantum theory. Thus quantum controlled measurements make the fundamental structure of the Hilbert space formalism accessible to direct experimental observation at arbitrary measurement strengths. 

\section{Quantum controlled measurement interactions}

Let us first consider a quantum controlled measurement of a $d$-level system, where the state of a single qubit decides whether a fully projective measurement of $\hat{A}$ is performed or not. In either case, there will be a measurement result $A_a$ associated with an eigenstate $\mid a \rangle$ of the observable $\hat{A}$. However, if no measurement was performed, the result will be completely random, with a probability of $1/d$ for each of the $d$ outcomes $a$. The effect of the quantum controlled measurement can then be described by a measurement operator $\hat{E}(a)$ that acts on the input states of system and control qubit according to
\begin{equation}
\hat{E}(a)= \frac{1}{\sqrt{d}}\hat{I}\otimes\mid 0 \rangle\langle 0 \mid + \mid a \rangle\langle a \mid \otimes \mid 1 \rangle\langle 1 \mid, 
\end{equation}
where $\hat{I}$ is the identity operator of the $d$-dimensional Hilbert space.

If the input state of the control qubit is an equal superposition of $\mid 0 \rangle$ and $\mid 1 \rangle$, all possible quantum superpositions of projective measurement and identity operation can be selected by an appropriate measurement on the control qubit output. The operation on the system associated with the successful post-selection of a specific output state $\mid \theta, \phi \rangle$ can be expressed as
\begin{eqnarray}
\label{eq:select}
\hat{S}_{\theta,\phi}(a,1) &=& \frac{1}{\sqrt{2}} \langle(\theta, \phi) \mid \hat{E}(a) \;\;(\mid 0 \rangle + \mid 1 \rangle)
\nonumber \\
&=& \frac{1}{\sqrt{2}} \left(\frac{1}{\sqrt{d}} \cos \theta \hat{I} + \mathrm{e}^{i \phi} \sin \theta \mid a \rangle\langle a \mid \right),
\end{eqnarray}
where $(a,1)$ indicates a measurement result of $a$ followed by a successful post-selection of the control qubit in the corresponding superposition of $\mid 0 \rangle$ and $\mid 1 \rangle$. The probability $P(1)$ of obtaining the desired control qubit output $\mid \theta, \phi \rangle$ is independent of the input state of the system and can be evaluated from the sum over all outcomes $a$,
\begin{equation}
\sum_a \hat{S}^\dagger_{\theta,\phi}(a,1)\hat{S}_{\theta,\phi}(a,1) = P(1) \hat{I},
\end{equation}
where
\begin{equation}
\label{eq:Pnorm}
P(1) = \frac{1}{2}\left(1+ \frac{2}{\sqrt{d}} \left(\sin(\theta) \cos(\theta)\right) \cos(\phi)\right).
\end{equation}
Thus a phase of $\phi=0$ indicates constructive interference between the measurement and the identity operation, while $\phi=\pi$ indicates destructive interference. 

To evaluate the measurement resolution, we can now consider the probability of obtaining the correct output result for an eigenstate input $\mid a \rangle$. This measurement fidelity is given by 
\begin{eqnarray}
\label{eq:fidelity}
F &=& \frac{|| \hat{S}_{\theta,\phi}(a,1) \mid a \rangle ||^2 }{P(1)}
\nonumber \\
&=& 1 - \frac{d-1}{2 d P(1)} \; (\cos(\theta))^2
\end{eqnarray}
In the limit of maximal measurement strength ($\theta=\pi/2$), this fidelity is $1$, indicating a fully resolved projective measurement. In the limit of zero measurement strength ($\theta=0$), the fidelity is $1/d$, corresponding to the fidelity of a random guess. For intermediate measurement strengths, the fidelity also depends on the phase $\phi$ of the control qubit, with maximal fidelities achieved at $\phi=0$, where measurement and identity operation interfere constructively. 

Next, we can take a look at the measurement back-action by considering how the measurement changes the statistics of an input state $\hat{\rho}_\mathrm{in}$ if we do not consider the specific measurement outcome. These changes are described by the mixed state output
\begin{eqnarray}
\label{eq:backact}
\hat{\rho}_{\mathrm{out}} &=& \frac{1}{P(1)} \sum_a \hat{S}_{\theta,\phi}(a,1) \hat{\rho}_{\mathrm{in}} \hat{S}^\dagger_{\theta,\phi}(a,1)
\nonumber \\
&=& \eta \hat{\rho}_{\mathrm{in}} + (1-\eta) \sum_a \mid a \rangle \langle a \mid \hat{\rho}_{\mathrm{in}} \mid a \rangle \langle a \mid,
\end{eqnarray}
where the back-action reduces all coherences between different eigenstates of $\hat{A}$ by a constant dephasing factor of $\eta<1$, given by
\begin{equation}
\label{eq:dephase}
\eta =  1 - \frac{(\sin(\theta))^2}{2 P(1)}.
\end{equation}
Complete dephasing occurs at maximal measurement strength ($\theta=\pi/2$), while no dephasing is observed in the weak limit ($\theta=0$). For intermediate measurements, dephasing effects can be minimized by choosing a control qubit phase of $\phi=0$, indicating that constructive interference between measurement and identity optimizes the trade-off between measurement errors and back-action. 

\section{Sequential measurement of two non-commuting observables}

Since the strength of the quantum controlled measurement is variable, it is now possible to explore the joint quantum statistics of two non-commuting observables $\hat{A}$ and $\hat{B}$ directly by first performing the controlled measurement of $\hat{A}$ and then performing a precise projective measurement of $\hat{B}$. In this sequential measurement, the measurement back-action of the quantum controlled measurement changes the statistics of $\hat{B}$ observed in the output, as described by the dephasing term in Eq.(\ref{eq:backact}). At maximal measurement strength, the ``collapse'' of the state into an eigenstate of $\hat{A}$ is reproduced, while the weak limit of zero measurement strength leaves the initial quantum state unchanged. In the intermediate regime, the measurement results in a non-trivial joint probability of the outcomes $a$ and $b$, corresponding to a joint measurement of the non-commuting observables $\hat{A}$ and $\hat{B}$.

The normalized joint probability obtained for a specific post-selected control qubit state is given by
\begin{eqnarray}
\label{eq:expjoint}
p(a,b|1) &=& \frac{1}{P(1)}  \langle b \mid \hat{S}_{\theta,\phi}(a,1) \; \hat{\rho}_{\mathrm{in}} \; \hat{S}^\dagger_{\theta,\phi}(a,1) \mid b \rangle
\nonumber 
\\[0.1cm] &=& \frac{1}{2 d P(1)} \bigg(
(\cos \theta)^2 \; \langle b \mid \hat{\rho}_{\mathrm{in}} \mid b \rangle
+ d (\sin \theta)^2|\langle b \mid a \rangle|^2 \;
\langle a \mid  \hat{\rho}_{\mathrm{in}}\mid a \rangle
\nonumber 
\\[0.1cm] && \hspace{1.5cm}
\; \; + \;  2 \sqrt{d} \sin \theta \cos \theta \;\; \mbox{Re}\left(\mathrm{e}^{i \phi} \langle b \mid a \rangle \langle a \mid \hat{\rho}_{\mathrm{in}}\mid b \rangle \right) \bigg).
\end{eqnarray}
It is possible to distinguish three separate contributions to this joint probability of $a$ and $b$. The first contribution originates from the identity operation and has equal probabilities for all outcomes $a$. As expected, this contribution describes the zero strength measurement at $\theta=0$ and produces the correct marginal probability of $\langle b \mid \hat{\rho}_{\mathrm{in}} \mid b \rangle$ for the outcomes $b$ of the final measurement. The second term originates from the projective measurement and produces the correct marginal probabilities of $\langle a \mid \hat{\rho}_{\mathrm{in}} \mid a \rangle$ for the outcomes $a$ of the quantum controlled measurement. The probabilities of the different final outcomes $b$ are then obtained by projection onto the eigenstates of $a$.

The third contribution originates from the superposition of identity operation and projective measurement and therefore represents the effects of quantum coherence in the quantum controlled measurement. Significantly, this contribution to the measurement outcomes reproduces the correct distributions of $a$ and $b$ for the unchanged input state $\hat{\rho}$ in the marginal probabilities. It therefore seems that the quantum coherence between identity and measurement projection represents an error free joint measurement of $a$ and $b$. However, the phase factor $\phi$ also plays an important role in determining the contributions to the joint probabilities. Specifically, the third term in the experimental joint probability $p(a,b|1)$ given in Eq.(\ref{eq:expjoint}) can be expressed in terms of a complex joint probability $\rho(a,b)$ obtained from the expectation values of an ordered product of the projections on $a$ and $b$,
\begin{eqnarray}
\label{eq:jointprob}
\rho(a,b) &=& \mbox{Tr} \left(\;\mid b \rangle \langle b \mid a \rangle \langle a \mid \hat{\rho}_{\mathrm{in}}\; \right)
\nonumber \\ &=&
\langle b \mid a \rangle \langle a \mid \hat{\rho}_{\mathrm{in}} \mid b \rangle.
\end{eqnarray}
This quantum mechanical analog of a joint probability of $a$ and $b$ has recently been discussed extensively in the context of weak measurements \cite{Joh07,Lun12,Hof12,Sal13,Bam13}. Specifically, it can be obtained by interpreting the complex weak value of the operator $\mid a \rangle\langle a \mid$ as conditional probability for the final measurement result $b$. The joint probability is then obtained by multiplying that weak value with the final outcome probability of $\langle b \mid \hat{\rho} \mid b \rangle$. 

Interestingly, the complex joint probability $\rho(a,b)$ is also a complete mathematical representation of the quantum state $\hat{\rho}$, a fact that was already pointed out in the early days of quantum mechanics \cite{McCoy32,Kir33,Dir45}. For this reason, weak measurements of $\rho(a,b)$ provide a particularly direct method of quantum state tomography, as discussed in several recent papers \cite{Lun12,Sal13,Bam13}. The present result shows that quantum controlled measurements may provide an alternative method of quantum tomography based on the complex joint probabilities of two non-commuting observables: if the effects of measurement errors and the back-action disturbance are properly accounted for, the intrinsic joint probability responsible for the correlations between $a$ and $b$ observed in the experimental data given by $p(a,b|1)$ can be identified with the appropriate phase component of the complex joint probability $\rho(a,b)$.

\section{Analysis of measurement errors and back-action}

Before proceeding to the reconstruction of the complex joint probability $\rho(a,b)$ from the experimental data, it may be useful to summarize the identification of the measurement and back-action errors in the experimental joint probability $p(a,b|1)$. For this purpose, Eq.(\ref{eq:expjoint}) can be reformulated by normalizing the three contributions associated with the identity operation, the projective measurement, and the quantum coherence between the two. The coefficients of the normalized contributions then correspond to quasi probability associated with the three processes,
\begin{eqnarray}
\label{eq:decomp}
p(a,b|1) &=& P_I \; \frac{1}{d} \langle b \mid \hat{\rho}_{\mathrm{in}} \mid b \rangle + P_M \; |\langle b \mid a \rangle|^2 \;
\langle a \mid  \hat{\rho}_{\mathrm{in}}\mid a \rangle
\nonumber \\ && \hspace{1.5cm}
+ P_C \; \frac{\mbox{Re}\left(\mathrm{e}^{i \phi} \rho(a,b)\right)}{\cos \phi}. 
\end{eqnarray}
Since the contributions are all normalized to one, the sum of the three quasi probabilities $P_I$, $P_M$ and $P_C$ will also add up to one. Comparison with Eq.(\ref{eq:expjoint}) shows that the quasi probabilities are given by
\begin{eqnarray}
P_I &=& \frac{1}{2 P(1)} (\cos\theta)^2
\nonumber \\
P_M &=& \frac{1}{2 P(1)} (\sin\theta)^2
\nonumber \\
P_C &=& \frac{1}{2 P(1)} \frac{2 \cos(\phi)}{\sqrt{d}} \sin(\theta)\cos(\theta).
\end{eqnarray}
Since measurement errors only originate from the random assignment of outcomes associates with $P_I$, the fidelity $F$ of the $\hat{A}$-measurement given in Eq.(\ref{eq:fidelity}) can now be expressed in terms of this error probability,
\begin{equation}
\label{eq:PIerror}
F = 1 - \frac{d-1}{d} P_I.
\end{equation}
Likewise, dephasing only originates from the projective measurements associated with $P_M$. Therefore, the dephasing factor $\eta$ given in Eq.(\ref{eq:dephase}) can be expressed as
\begin{equation}
\label{eq:PMerror}
\eta = 1 - P_M.
\end{equation}
These results confirm that the quasi probabilities $P_I$ and $P_M$ correspond to the relative frequencies of measurement errors and back-action disturbance, respectively, while the contribution of the coherence given by $P_C$ contributes to neither errors nor disturbance and hence corresponds to a precise back-action free measurement. 

From the viewpoint of measurement theory, it seems obvious that the most precise measurement of $a$ and $b$ is obtained at $\phi=0$, where the trade-off between measurement errors and back-action disturbance is optimal. This means that the most direct evidence for intrinsic correlations between $a$ and $b$ is obtained when there is constructive interference between the identity operation and the projective measurement in the quantum controlled measurement of $a$. The intrinsic joint probability associated with the error free contribution of probability $P_C$ is then given by the real part of the complex joint probability $\rho(a,b)$, consistent with the results obtained in weak measurements, where the small shifts of the meter also indicate only the real parts of the weak values. However, it is important to note that there is additional information in the imaginary part of the joint probability, making it desirable to consider a complete reconstruction of the complex joint probability from the measurement data obtained at different control qubit phases $\phi$.

\section{Reconstruction of complex joint probabilities}

If the error statistics of a measurement are well understood, the intrinsic probabilities of the input state can be derived from the experimentally determined joint probability by a deconvolution of the errors \cite{Suz12}. In the present case, we can apply the error model of Eq.(\ref{eq:decomp}) to subtract the statistical background associated with measurement errors ($P_I$) and with back-action disturbance ($P_M$). The intrinsic joint probability is then identified with the normalized phase component of the complex joint probability $\rho(a,b)$,
\begin{eqnarray}
\label{eq:reconst}
\lefteqn{\frac{\mbox{Re}\left(\mathrm{e}^{i \phi} \rho(a,b)\right)}{\cos\phi} =}
\nonumber \\ &&
 \frac{1}{P_C} \left(p(a,b|1) -  P_I \; \frac{1}{d} \langle b \mid \hat{\rho}_{\mathrm{in}} \mid b \rangle - P_M \; |\langle b \mid a \rangle|^2 \;
\langle a \mid  \hat{\rho}_{\mathrm{in}}\mid a \rangle\right).
\end{eqnarray}
If the directly accessible probabilities of $a$ and $b$ in the quantum state $\rho$ are known from separate measurements, this simple background subtraction is sufficient for the reconstruction of the intrinsic joint probability of $a$ and $b$. Alternatively, the probabilities can also be obtained by applying the same kind of background subtraction to the marginal probabilities of $p(a,b|1)$,
\begin{eqnarray}
\label{eq:marginal}
\langle a \mid \hat{\rho} \mid a \rangle &=& \frac{1}{1-P_I}\left(\left(\sum_b p(a,b|1) \right) - \frac{1}{d} P_I \right)
\nonumber \\
\langle b \mid \hat{\rho} \mid b \rangle &=& \frac{1}{1-P_M}\sum_a \left( p(a,b|1) - P_M |\langle b \mid a \rangle|^2 \langle a \mid \hat{\rho} \mid a\rangle \right).
\end{eqnarray}
Note that this procedure involves only positive probabilities and has a clear and straightforward classical interpretation. Nevertheless, the application of this procedure to the joint probabilities of non-commuting observables results in a non-positive intrinsic joint probability that actually depends not only on the input state, but also on the phase $\phi$ of the control qubit. This dependence on a quantum mechanical phase can be represented in terms of a linear combination of the real and the imaginary parts of the complex joint probability $\rho(a,b)$,
\begin{equation}
\label{eq:intrinsic}
\frac{\mbox{Re}\left(\mathrm{e}^{i \phi} \rho(a,b)\right)}{\cos\phi} = \mbox{Re}\left(\rho(a,b)\right) - \tan \phi \; \mbox{Im}\left(\rho(a,b)\right).
\end{equation}
Thus the change of phase $\phi$ redistributes the measurement outcomes $(a,b)$ in a way that leaves all marginals unchanged, while modifying the correlations between $a$ and $b$ observed in the experimental data. 

In principle, the complete complex probability - and therefore the complete quantum state $\hat{\rho}$ - can be reconstructed from only two settings of $\phi$. In particular, Eq.(\ref{eq:intrinsic}) suggests that the settings with $\tan \phi = \pm 1$ would provide optimal balance between sensitivity to the real part and sensitivity to the imaginary part. Alternatively, a more direct evaluation of the data can be obtained by exploiting the phase dependence of the experimental probabilities $p(a,b|\phi)$ obtained by continuously varying $\phi$ over a whole period of $2 \pi$. The complex joint probability can then be observed as a control qubit visibility, where the phase dependence of the output probability $P(1)=P(\phi)$ should be included in the evaluation as follows,
\begin{equation}
\label{eq:phase}
\frac{1}{2\pi} \int p(a,b|\phi) P(\phi) \exp(- i \phi) = \frac{\sin\theta\cos\theta}{\sqrt{d}}.
\end{equation}
In the limit of $\theta \to 0$, the quantum controlled measurement is a weak measurement and the method of evaluation given by Eq.(\ref{eq:phase}) is equivalent to the one used in previous experiments \cite{Lun12,Sal13,Bam13}. By using a quantum controlled measurement, the same result can be obtained at any measurement strength, with a considerable increase in the signal-to-noise ratio. Optimal results will be obtained at $\theta=\pi/4$, where the visibility of the signal is only limited by the dimensionality $d$ of the Hilbert space. 

\section{Complex probabilities and non-commutativity}

In the weak measurement limit, the imaginary part of the weak value is usually associated with the response of the system to dynamics generated by the target observable \cite{Hof11,Dre12b}. In the quantum controlled measurement, the imaginary part of the joint probability appears as an independent component of the intrinsic joint probability. Thus the phase of the control qubit establishes a more direct connection between the dynamical structure of quantum mechanics and the joint measurement statistics of non-commuting observables. 

In a purely statistical interpretation of the measurement data, it seems strange that different intrinsic probabilities can be obtained depending on the phase selected for the control qubit, especially since the marginal distribution of $a$ and $b$ is unchanged by the addition, no matter what the actual input state is. Effectively, Eq.(\ref{eq:intrinsic}) describes an ambiguity in the intrinsic correlations between $\hat{A}$ and $\hat{B}$ described by the quantum state $\hat{\rho}$. To understand the nature of this ambiguity, it is useful to consider the complex correlations obtained by averaging the products of the eigenvalues $A_a$ and $B_b$ over the complex joint probability,
\begin{equation}
\label{eq:comcorr}
\sum_{a,b} A_a B_b \rho(a,b) = \mbox{Tr}\left(\hat{B} \hat{A} \hat{\rho} \right).
\end{equation}
In general, $\rho(a,b)$ is the joint probability of $a$ and $b$ for which correlations of all orders are given by the expectation values of the well-ordered operator products, where all operators $\hat{A}$ are placed on the right and all operators $\hat{B}$ are placed on the left. Note that this is a fundamental feature of the complex joint probability $\rho(a,b)$ that has actually been used by Dirac to derive it from purely formal mathematical criteria in \cite{Dir45}. Due to this selection of a particular operator ordering, the non-commutativity of $\hat{A}$ and $\hat{B}$ results in imaginary probabilities. In particular, the imaginary part of the correlation between $\hat{A}$ and $\hat{B}$ is given by the expectation value of the commutation relation,
\begin{equation}
\label{eq:icorr}
\sum_{a,b} A_a B_b \mbox{Im}\left(\rho(a,b)\right) = 
\frac{i}{2} \mbox{Tr}\left(\left[ \hat{A}; \hat{B} \right] \hat{\rho} \right).
\end{equation}
Thus the imaginary parts of complex joint probabilities represent an alternative expression of non-commutativity. In the quantum controlled measurement, the selection of a specific control qubit phase converts these imaginary correlations into real correlations, allowing a direct measurement of the commutation relation averages for all operator pairs with eigenstates of $\mid a \rangle$ and $\mid b \rangle$. 

Although imaginary correlations do not exist in classical physics, the role of commutation relations can be identified with the role of Poisson brackets in the classical dynamics of a system. Thus, complex joint probabilities may provide a more detailed description of the fundamental relation between dynamics and statistics suggested by the operator algebra of quantum mechanics \cite{Hof11,Hof13}. In the quantum controlled measurement, the essential link between measurement statistics and dynamics is established by the measurement operators $\hat{S}_{\theta,\phi}(a,1)$ shown in Eq.(\ref{eq:select}). For values of $\phi$ other than zero or $\pi$, these operators are not hermitian and therefore include the effects of a unitary transformation. Importantly, the relation between the measurement outcome $a$ and the unitary operation is established by the exclusive application of a phase shift to the projector describing the single measurement outcome $a$. In the sequential measurement of $a$ and $b$, the total effect of the unitaries in $a$ on the marginal distribution of $b$ cancels out, so that only the correlations between $a$ and $b$ are modified by the unitary operations associated with the non-hermitian operators $\hat{S}_{\theta,\phi}(a,1)$.

\section{Quantum coherence and the logical ``AND''}

In the analysis above, the essential element is the statistical decomposition of the experimentally observable measurement probabilities into contributions from two distinct processes and a third contribution associated with the quantum coherence between the two processes. As the analysis shows, the non-positive contribution associated with quantum coherence describes an uncertainty-free contribution to the measurement process, corresponding directly to a simultaneous assignment of eigenvalues to the non-commuting properties $\hat{A}$ and $\hat{B}$. The statistical weight of this simultaneous assignment in the initial quantum state $\hat{\rho}$ is represented by the complex joint probability $\rho(a,b)$. As shown in Eq.(\ref{eq:jointprob}), the operator that represents this joint assignment in the Hilbert space formalism is the product of the projection operators $\mid a \rangle\langle a \mid$ and $\mid b \rangle\langle b \mid$.

Fundamentally, the problem of measurement uncertainty arises because quantum mechanics does not permit any consistent simultaneous assignment of outcomes to the measurements of the non-commuting observables $\hat{A}$ and $\hat{B}$. However, this should not be confused with mere ignorance: the formalism does permit very specific statements about the relations between the outcomes. In particular, the product of the projection operators $\mid a \rangle\langle a \mid$ and $\mid b \rangle\langle b \mid$ is the quantum mechanical equivalent of a product of truth values, and therefore corresponds to the logical ``AND'' for the statements $a$ and $b$. The fact that this operator is non-hermitian and cannot be reduced to a more precise projection onto a joint reality is a strong indication that the quantum formalism cannot be reconciled with any assignment of joint realities for $a$ and $b$. Oppositely, the experimental evidence that can be obtained in both weak and quantum controlled measurements clearly points to complex valued joint probabilities for measurements that cannot be performed jointly. In the light of this evidence, the reason for uncertainty limits in both measurement and state preparation appears to be that a sufficient amount of statistical noise is required to ``cover up'' the negative and the imaginary probabilities associated with the error-free contribution associated with the quantum coherence that describes the quantum mechanical ``AND'' in the operator formalism. 

\section{Conclusions}

In a sequential measurement of two non-commuting observables, the accuracy of the results is limited by the measurement resolution and the back-action disturbance of the initial measurement. However, it may still be possible to observe the non-classical correlations between the observables in the experimentally obtained statistics if the statistical effects of resolution errors and back-action disturbance can be explained by a sufficiently simple model. Quantum controlled measurements are a particularly promising candidate for this kind of analysis, since the quantum superposition of identity operation and fully projective measurement limits the possible statistical patterns to only three: an unbiased measurement error associated with the random assignment of an outcome in the absence of a measurement interaction, a precise projective measurement with the corresponding back-action disturbance, and a contribution from the coherence between the identity operation and the projective measurement. Significantly, the coherence between identity and measurement projection does not contribute anything to the measurement errors or the back-action disturbance of the measurement, and thus appears to represent an uncertainty free measurement that returns the correct value of $a$ without any reduction in the quantum coherence of the state. 

The data obtained in a quantum controlled measurement of $a$ followed by a final measurement of $b$ can be explained as a statistical mixture of measurement error, back-action disturbance, and uncertainty free measurement. By subtracting the statistical background associated with the measurement uncertainties, it is possible to identify the intrinsic joint probability of $a$ and $b$ in the experimental results. Due to the dependence of the results on the phase of the superposition determined by the control qubit measurement, this intrinsic probability has two independent components. These components can be identified with the real and the imaginary part of the complex joint probability obtained from the expectation value of the projector product $\mid b \rangle\langle b \mid a \rangle \langle a \mid$, a quasi probability that has been known as a quantum analog of classical phase space distributions from the early days of quantum mechanics, and has recently been observed directly in weak measurements \cite{McCoy32,Kir33,Dir45,Joh07,Lun12,Hof12,Sal13,Bam13}. Its direct observation in quantum controlled measurements allows a complete reconstruction of the input state from the experimental data obtained in the sequential measurement of $a$ and $b$. Since the quantum controlled measurement can be performed at any measurement strength, this approach can achieve much better signal-to-noise ratios than in the corresponding measurements carried out in the weak measurement limit \cite{Lun12,Sal13,Bam13}. 

Once the statistical background associated with measurement uncertainties is subtracted, the experimental data obtained in quantum controlled measurements provides a detailed characterization of the non-classical correlations between non-commuting observables. In particular, the quantum phase of the control qubit output converts the imaginary part of the complex joint probability into a real contribution to the experimental joint probability. Thus characteristic quantum features such as the expectation value of the commutation relation of $\hat{A}$ and $\hat{B}$ can be observed directly in the experimentally accessible measurement statistics. Quantum controlled measurements may thus be a powerful tool for the exploration of the quantum correlations described by non-commuting operators. 

\section*{Acknowledgment}
This work was supported by JSPS KAKENHI Grant Number 24540427.

\end{document}